\pdfoutput=1
\documentclass[aps,floats,twocolumn,nofootinbib]{revtex4-2}

\usepackage{amsfonts,amsmath,amssymb,ascmac,bm,tensor}
\usepackage{fnpct} 
\usepackage{comment}
\usepackage{ifpdf}
\usepackage{slashed}
\usepackage{esint}
\usepackage{color}
\usepackage[mathscr]{eucal}
\usepackage[utf8]{inputenc}
\usepackage{physics}
\usepackage{cancel}
\usepackage{soul}
\usepackage{simpler-wick}

\ifpdf
  \usepackage{graphicx}     
  \usepackage[bookmarksopen,colorlinks=true,linkcolor=bblue,citecolor=bblue,urlcolor=ppink]{hyperref}
\else     
\fi

\hypersetup{
           breaklinks=true,   
           colorlinks=true,   
           linkcolor=bblue,
           citecolor=bblue,
           urlcolor=ppink
        }

\definecolor{red}{rgb}{1,0,0}
\definecolor{darkred}{rgb}{0.6,0,0}
\definecolor{darkgreen}{rgb}{0.992447,0.623778,0.034597}
\definecolor{ppink}{rgb}{1,0.4,0.4} 
\definecolor{bblue}{rgb}{0.284602,0.317763,0.963947}
\definecolor{purple}{rgb}{0.5 ,0, 0.7}

\newcommand{\fo}{{(1)}}
\renewcommand{\so}{{(2)}}
\newcommand{\tho}{{(3)}}
\newcommand{\foo}{{(4)}}
\newcommand{\no}{{(n)}}

\newcommand{\tre}{\text{tr} }
\newcommand{\vx}{\text{vx}}

\newcommand{\bulk}{\text{bulk}}

\newcommand{\sbra}[1]{\langle #1 \rangle}

\newcommand{\inte}{\text{int}}

\newcommand{\Pl}{\text{Pl} }
\newcommand{\ee}{\text{e}}

\newcommand{\uv}{\text{UV}}
\newcommand{\ir}{\text{IR}}

\newcommand{\pp}{\text{p}}

\newcommand{\bfk}{\mathbf{k}}

\newcommand{\bfq}{\mathbf{q}}
\newcommand{\bfx}{\mathbf{x}}
\newcommand{\bfy}{\mathbf{y}}

\renewcommand{\Re}{\text{Re}}
\renewcommand{\Im}{\text{Im}}

\makeatletter
\newcommand\footnoteref[1]{\protected@xdef\@thefnmark{\ref{#1}}\@footnotemark}
\makeatother

\allowdisplaybreaks[1]

\begin{document}


\title{
Role of the counterterms in the conservation of superhorizon curvature perturbations at one loop
}

\author{Keisuke Inomata}
\affiliation{William H. Miller III Department of Physics and Astronomy, Johns Hopkins University, 3400 N. Charles Street, Baltimore, Maryland, 21218, USA}

\begin{abstract} 
\noindent
Recently, several papers have claimed that superhorizon curvature perturbations are not conserved at the one-loop level in single-field inflation models if there is a transient ultra-slow-roll period.
In this work, we point out that the contributions from the counterterms were overlooked in the recent papers.
We show that the counterterm contributions play a crucial role in canceling the one-loop power spectrum of superhorizon curvature perturbations in the comoving gauge.
\end{abstract}

\date{\today}
\maketitle

\section{Introduction} 
Cosmological perturbations serve as a valuable probe of the early Universe. They were generated in the early Universe and have been observed through the cosmic microwave background anisotropies and the large-scale structure. The amplitude of these perturbations is often characterized using curvature perturbations, as they are expected to remain constant (or conserved) once they exit the horizon in single-field inflation models~\cite{Bardeen:1980kt,Kodama:1984ziu,Lyth:2004gb,Senatore:2009cf,Senatore:2012nq,Pimentel:2012tw}.

However, Ref.~\cite{Kristiano:2022maq} recently claimed that, if there is a transient ultra-slow-roll (USR) period during the inflation, the superhorizon curvature perturbations are not conserved at the one-loop level even in single-field models.\footnote{See also Ref.~\cite{Cheng:2021lif} for an earlier work claiming the non-conservation of the superhorizon curvature perturbations with the use of the Hartree factorization.} 
Specifically, the claim can be summarized as follows: 
Suppose the USR period occurs only within $\tau_i < \tau < \tau_e$, where $\tau$ is the comoving time, while the other periods are in the slow-roll (SR) phase. Then, the one-loop curvature power spectrum is given by 
\begin{align}
  \lim_{q\to 0}\mathcal P^{1\text{-}\text{loop}}_{\zeta}(q) = \begin{cases}
  0 & (\tau < \tau_i) \\
  c\, \mathcal P_{\zeta,\tre}(q) \displaystyle\int_{k_\ir} \dd \ln k\, \mathcal P_{\zeta,\tre}(k) & (\tau > \tau_e)
  \end{cases},
\end{align}
where $\mathcal P_{\zeta,\tre}$ is the tree-level power spectrum, $c$ is independent of $q$, and $k_\ir$ is some IR cutoff ($k_\ir \gg q$).
The claim of Ref.~\cite{Kristiano:2022maq} is that $c$ can be $\gtrsim 1$ if a sharp transition between the USR and the SR is considered.

This claim has sparked controversy~\cite{Riotto:2023hoz,Choudhury:2023vuj,Kristiano:2023scm,Riotto:2023gpm,Firouzjahi:2023aum,Motohashi:2023syh,Firouzjahi:2023ahg,Franciolini:2023agm,Tasinato:2023ukp,Cheng:2023ikq,Maity:2023qzw,Firouzjahi:2023bkt,Davies:2023hhn,Iacconi:2023ggt,Saburov:2024und,Ballesteros:2024zdp,Kristiano:2024vst,Kristiano:2024ngc,Sheikhahmadi:2024peu,Frolovsky:2025qre,Fumagalli:2023hpa,Tada:2023rgp,Inomata:2024lud,Kawaguchi:2024rsv,Fumagalli:2024jzz,Inomata:2025bqw,Fang:2025vhi}.
Some papers accept that $c$ can be nonzero, while others argue that $c = 0$.
A nonzero $c$ implies the non-conservation of curvature perturbations in the superhorizon limit, which appears to be inconsistent with the separate universe picture~\cite{Lyth:2004gb}.
In this work, we focus on the question of whether $c = 0$ or not.

The authors in Refs.~\cite{Fumagalli:2023hpa,Tada:2023rgp,Inomata:2024lud,Kawaguchi:2024rsv,Fumagalli:2024jzz,Inomata:2025bqw,Fang:2025vhi} have shown $c=0$ using different approaches.
Refs.~\cite{Fumagalli:2023hpa,Tada:2023rgp,Fumagalli:2024jzz} use the comoving gauge and take into account the boundary terms. 
Ref.~\cite{Kawaguchi:2024rsv} uses the path-integral approach. 
Refs.~\cite{Inomata:2024lud,Inomata:2025bqw} (our previous papers) use the spatially-flat gauge. 
Ref.~\cite{Fang:2025vhi} utilizes the gauge transformation. 
This raises a question:
If these works are correct, what was overlooked in the original study~\cite{Kristiano:2022maq} and its follow-up~\cite{Kristiano:2024vst}?
The original works take the comoving gauge but use the field redefinition of curvature perturbations to remove the boundary term contributions. 
This makes it difficult to directly compare their results with those in Refs.~\cite{Tada:2023rgp,Fumagalli:2023hpa,Fumagalli:2024jzz}, even though they use the same gauge.
The main goal of this work is to address this question.

To this end, we connect the results in the spatially-flat gauge~\cite{Inomata:2024lud,Inomata:2025bqw} to those in the comoving gauge. 
Specifically, we use the fact that the redefined curvature perturbation is proportional to the inflaton fluctuation in the spatially-flat gauge. 
A similar approach has been done in Ref.~\cite{Fang:2025vhi}.
The main difference is that we take into account the counterterms and impose the zero-tadpole condition $\expval{\zeta_n} = 0$, where $\zeta_n$ represents the redefined curvature perturbation.
Under this condition, backreaction effects are absent, and unlike Ref.~\cite{Fang:2025vhi}, we do not need to introduce backreaction terms in our analysis. 
This setup aligns more closely with previous works, as most of them either explicitly or implicitly impose the zero-tadpole condition, effectively neglecting tadpole contributions.

Then, we will see that the counterterms play an important role in the cancellation of the one-loop power spectrum of the superhorizon curvature perturbations. 
This counterterm contribution was overlooked in the original studies~\cite{Kristiano:2022maq,Kristiano:2024vst}, as well as in many subsequent works.
Notably, the importance of the counterterm contribution for the curvature conservation has been emphasized without using the field redefinition in Refs.~\cite{Pimentel:2012tw,Fumagalli:2024jzz}.
We will also discuss the case without the zero-tadpole condition in the last section.

Throughout this work, we consider the de Sitter limit ($\epsilon (\equiv -\dot H/H^2) \to 0$) and neglect the higher-order terms in $\epsilon$ for simplicity.

\section{Setup}
We use the expressions of the Hamiltonian in the spatially-flat gauge to obtain those in the comoving gauge with the redefined curvature perturbations. 
The free part of the Hamiltonian in the spatially-flat gauge is given by~\cite{Inomata:2025bqw}
\begin{align}
  \label{eq:free_h}
  \mathcal H_0 = \frac{1}{2a^2} \left[(\delta \phi')^2 + (\partial_i \delta \phi)^2\right] + \frac{1}{2} V_\so(\bar \phi) \delta \phi^2,
\end{align}
where $V_\no \equiv \dd^n V(\phi)/\dd \phi^n$ and $\bar \phi$ is the background inflaton value.
We assume without loss of generality that the inflaton background velocity is positive, $\dot{\bar \phi} > 0$.
The interaction Hamiltonian that is relevant to the one-loop contributions can be split as~\cite{Inomata:2025bqw} 
\begin{align}
  \mathcal H_\inte = \mathcal H_{\inte, 3} + \mathcal H_{\inte, 4} + \mathcal H_c,
\end{align}
where 
\begin{align}
  \label{eq:int_h0}
  \mathcal H_{\inte,3} &= \frac{1}{6} V_\tho \delta \phi^3 , \ \mathcal H_{\inte,4} = \frac{1}{24} V_\foo \delta \phi^4, \\
  \label{eq:int_hc}  
  \mathcal H_c &= V_{c,\fo} \delta \phi + \frac{1}{2}V_{c,\so} \delta \phi^2.
\end{align}
We note that $\mathcal H_c$ is the counterterm contribution, which is not carefully taken into account in previous works.
We stress that $V_{c,\so} = \dd V_{c,\fo}/\dd \phi$ must hold due to the continuity of the inflaton potential~\cite{Inomata:2025bqw}.
In this work, we consider the transition from SR $\to$ non-SR $\to$ SR.
Specifically, we assume that $V_{\no >2}(\phi) \neq 0$ only when $\phi_i < \phi < \phi_e$. 
We here define $\tau_i$ and $\tau_e$ as $\bar\phi(\tau_{i(e)}) = \phi_{i(e)}$.
In addition, we assume that the small-scale curvature perturbations are enhanced during the non-SR period, which can be a USR period or a parametric resonance period~\cite{Inomata:2022yte}.

We here change the variable: 
\begin{align}
  \label{eq:zeta_n}
  \zeta_n \equiv - \frac{H}{\dot{\bar \phi}} \delta \phi = -\frac{\delta \phi}{\sqrt{2\epsilon}M_\Pl}. 
\end{align}
$\zeta_n$ is related to the curvature perturbation in the comoving gauge as 
\begin{align}
  \zeta = \zeta_n - f(\zeta_n),
\end{align}
where the expression of $f(\zeta_n)$ up to $\mathcal O(\zeta_n^3)$ can be seen in Ref.~\cite{Jarnhus:2007ia}.\footnote{
In the original studies~\cite{Kristiano:2022maq,Kristiano:2024vst}, the field redefinition is explicitly done only up to $\mathcal O(\zeta_n^2)$. 
Accordingly, the $\mathcal O(\zeta_n^4)$ Hamiltonian is not considered except for the induced quartic Hamiltonian (Eq.~(\ref{eq:bulk_4i})). }
This $\zeta_n$ is the redefined curvature perturbation, introduced in Ref.~\cite{Maldacena:2002vr} and used in many works.
During a SR period, $f(\zeta_n)$ becomes negligible and we have $\zeta \simeq \zeta_n$.
Since we are interested in whether the superhorizon curvature perturbations before and after the non-SR are the same, we can use $\zeta_n$ instead of $\zeta$ throughout this work. 
For brevity, we express $\zeta_n$ as $\zeta$ in the following unless otherwise noted.

Substituting Eq.~(\ref{eq:zeta_n}) into Eqs.~(\ref{eq:free_h}), (\ref{eq:int_h0}), and (\ref{eq:int_hc}), we obtain 
\begin{align}
  \mathcal H_0 &= \frac{1}{2a^2} b^2 \left[ \zeta^{\prime 2} + (\partial_i \zeta)^2\right], \\
  \label{eq:int_h0_z}
  \mathcal H_{\inte,3} &= \frac{1}{6} V_\tho b^3 \zeta^3, \ \mathcal H_{\inte,4} = \frac{1}{24} V_\foo b^4 \zeta^4, \\ 
  \mathcal H_c &= b V_{c,\fo} \zeta + \frac{b^2}{2}V_{c,\so} \zeta^2,
\end{align}
where $b(\tau) \equiv -\dot{\bar \phi}/H = -\sqrt{2\epsilon(\tau)}M_\Pl$.
After some calculation, we find the following relation in the de Sitter limit:
\begin{align}
\label{eq:v_tho}
&V_\tho = - a^{-4} (2\epsilon)^{-3/2} M_\Pl^{-1} (a^2 \epsilon \eta')', \\
&V_\foo = \frac{1}{a\dot{\bar \phi}} \frac{\dd V_\tho}{\dd \tau} = - \frac{(a^{-4} (2\epsilon)^{-3/2} (a^2 \epsilon \eta')')'}{\sqrt{2\epsilon}a H M_\Pl^2 },
\end{align}
where $\eta \equiv \dot \epsilon/(\epsilon H)$.
These are the Hamiltonian in the comoving gauge with the redefined curvature perturbations.
In Appendix~\ref{app:comp}, we see that substituting Eq.~(\ref{eq:v_tho}) into Eq.~(\ref{eq:int_h0_z}) indeed reproduces the calculation of Ref.~\cite{Kristiano:2024vst}.

In this work, we use the in-in formalism to calculate the power spectrum or the tadpole~\cite{Weinberg:2005vy}:
\begin{align}
  \expval{Q(\tau)} = \bra{0} (T \ee^{-i \int^\tau_{-\infty} \dd \tau' H_\text{int}})^\dagger Q^I(\tau) (T \ee^{-i \int^\tau_{-\infty} \dd \tau' H_\text{int}}) \ket{0},
  \label{eq:in_in}
\end{align}
where $Q$ is $\zeta(\bfx)$ or $\zeta_{\bfq} \zeta_{\bfq'}$, the superscript $I$ represents the operator in the interaction picture, $\ket{0}$ is the non-interacting vacuum, and 
\begin{align}
  H_{\cdots} \equiv \int \dd^3 x\,\, a^4 \mathcal H_{\cdots},
  \label{eq:int_hamil}
\end{align}
with $\cdots$ being '\inte', `$\inte,3$', `$\inte,4$', or `$c$'.
\begin{widetext}
Specifically, the power spectrum is given by 
\begin{align}
  &\expval{\zeta_{\bfq}(\tau) \zeta_{\bfq'}(\tau)} = \vev{\zeta^I_{\bfq}(\tau) \zeta^I_{\bfq'}(\tau) } + 2\, \Im\left[\int^\tau_{\tau_i} \dd \tau'\vev{ \zeta^I_{\bfq}(\tau) \zeta^I_{\bfq'}(\tau) H_{\text{int},4}(\tau')}\right] \nonumber \\
  &+ 2\, \Re\left[ \int^\tau_{\tau_i}\dd \tau' \int^{\tau'}_{\tau_i}\dd \tau'' \vev{ \left(H_{\text{int},3}(\tau') \zeta^I_{\bfq}(\tau) \zeta^I_{\bfq'}(\tau) - \zeta^I_{\bfq}(\tau) \zeta^I_{\bfq'}(\tau) H_{\inte,3}(\tau') \right) H_{\text{int},3}(\tau'')}\right] \nonumber \\
  &+ 2\, \Im\left[\int^\tau_{\tau_i} \dd \tau'\vev{ \zeta^I_{\bfq}(\tau) \zeta^I_{\bfq'}(\tau) H_{c}(\tau')}\right] \nonumber \\
  &= (2\pi)^3 \delta(\bfq + \bfq') \frac{2\pi^2}{q^3} \mathcal P_{\zeta}(q,\tau),
  \label{eq:two_vx_one_loop}
\end{align}
where we have used the fact that $V_{\no (>2)}(\tau) = 0$ in $\tau < \tau_i$.
\end{widetext}
We have expanded the curvature perturbations with the Fourier modes as 
\begin{align}
  \zeta^I(\bfx,\tau) &= \int \frac{\dd^3 k}{(2\pi)^3} \ee^{i \bfk \cdot \bfx} \zeta^I_{\bfk}(\tau)\nonumber \\
  &= \int \frac{\dd^3 k}{(2\pi)^3} \ee^{i \bfk \cdot \bfx} \left[ \zeta_k(\tau) a(\bfk) +  \zeta^{*}_k(\tau) a^{\dagger}(-\bfk) \right],
\end{align}
where $[a(\bfk), a(\bfk')] = [a^\dagger(\bfk), a^\dagger(\bfk')] = 0$ and $[a(\bfk), a^\dagger(-\bfk')] = (2\pi)^3 \delta(\bfk + \bfk')$.
For convenience, we split $\mathcal P_\zeta$ as\footnote{This split is slightly different from that in Ref.~\cite{Inomata:2025bqw}, where the counterterm contributions are included in $\mathcal P_{1\vx}$ and $\mathcal P_{2\vx}$.}
\begin{align}
  \mathcal P_\zeta = \mathcal P_{\zeta, \tre} + \mathcal P_{\zeta, 1\vx} + \mathcal P_{\zeta, 2\vx} + \mathcal P_{\zeta, c}, 
\end{align}
where $\mathcal P_{\zeta,\tre}$ is the tree-level contribution, $\mathcal P_{\zeta,1\vx}$ is from $H_{\inte,4}$ (the second term in the first line in Eq.~(\ref{eq:two_vx_one_loop})), $\mathcal P_{\zeta,2\vx}$ is from $H_{\inte,3}$ (the second line), and $\mathcal P_{\zeta,c}$ is from $H_{c}$ (the third line).
Note that there is no tadpole contribution because we impose $\sbra{\zeta} = 0$ by tuning the counterterm, which we will explicitly calculate later.

\section{Power spectrum}
The tree-level power spectrum is given by 
\begin{align}
  \mathcal P_{\zeta,\tre}(q,\tau) = \frac{q^3}{2\pi^2}|\zeta_q(\tau)|^2.
\end{align}
In $|q\tau_i| \ll 1$, we find $\mathcal P_{\zeta,\tre}(q,\tau) = \mathcal P_{\zeta,\tre}(q) (=\text{const.})$.
The one-vertex contribution is given by 
\begin{align}
  \label{eq:p_1vx}
  &\mathcal P_{\zeta,1\vx}(q,\tau) = \frac{q^3}{\pi^2} \int^{\tau}_{\tau_i} \dd \tau' a^4(\tau') b^4(\tau') V_\foo(\tau') \nonumber \\
  &\quad \times \Im[\zeta_q(\tau) \zeta^*_q(\tau')] \Re[\zeta_p(\tau) \zeta^*_p(\tau')] \sigma^2_\zeta(\tau'),
\end{align}
where $V_\no(\tau) \equiv V_\no(\bar \phi(\tau))$ and 
\begin{align}
  \sigma^2_\zeta(\tau) &\equiv \vev{(\zeta^I(\bfx, \tau))^2} = \int^{k_\uv \frac{a(\tau)}{a_i}}_{k_\ir} \dd \ln k \frac{k^3}{2\pi^2} |\zeta_k(\tau)|^2.
\end{align}
We have here introduced the physical UV cutoff scale and the comoving IR cutoff scale~\cite{Senatore:2009cf,Giddings:2010nc,Byrnes:2010yc,Senatore:2012nq}. 
The two-vertex contribution is given by 
\begin{widetext}
\begin{align}
  \mathcal P_{\zeta, 2\vx}(q,\tau) 
  &= -\frac{q^3}{\pi^2} \int^\tau_{\tau_i} \dd \tau' \int^{\tau'}_{\tau_i} \dd \tau'' a^4(\tau')  a^4(\tau'') b^3(\tau') b^3(\tau'') V_\tho(\tau') V_\tho(\tau'') \Re\left[ [\zeta_q(\tau) \zeta_q^*(\tau')- \zeta^*_q(\tau) \zeta_q(\tau')] 
   \zeta_q(\tau) \zeta_q^*(\tau'') \phantom{\int \frac{\dd^3 k}{(2\pi)^3}} \right. \nonumber \\
  &\qquad \qquad \qquad \left. \times \int \frac{\dd^3 k}{(2\pi)^3} u_k(\tau')\zeta^*_k(\tau'')u_{|\bfq - \bfk|}(\tau')\zeta^*_{|\bfq - \bfk|}(\tau'') \right] \nonumber \\
  &= \mathcal P^a_{\zeta, 2\vx}(q,\tau) + \mathcal P^b_{\zeta, 2\vx}(q,\tau),
  \label{eq:2vx}
\end{align}
where $\mathcal P^a_{\zeta, 2\vx}$ and $\mathcal P^b_{\zeta, 2\vx}$ are 
\begin{align}
  \label{eq:p_a}
  \mathcal P^a_{\zeta, 2\vx}(q,\tau) &= \frac{2q^3}{\pi^2}  \int^\tau_{\tau_i} \dd \tau' \int^{\tau'}_{\tau_i} \dd \tau'' a^4(\tau')  a^4(\tau'') b^3(\tau') b^3(\tau'') V_\tho(\tau') V_\tho(\tau'') \Im[\zeta_q(\tau) \zeta_q^*(\tau')] \Im[\zeta_q(\tau) \zeta_q^*(\tau'')] \nonumber \\
  &\qquad \qquad \qquad \times \int \frac{\dd^3 k}{(2\pi)^3} \Re[\zeta_k(\tau')\zeta^*_k(\tau'')\zeta_{|\bfq - \bfk|}(\tau')\zeta^*_{|\bfq - \bfk|}(\tau'')], \\
  \label{eq:p_b}  
  \mathcal P^b_{\zeta, 2\vx}(q,\tau) &= \frac{2q^3}{\pi^2}  \int^\tau_{\tau_i} \dd \tau' \int^{\tau'}_{\tau_i} \dd \tau'' a^4(\tau')  a^4(\tau'') b^3(\tau') b^3(\tau'') V_\tho(\tau') V_\tho(\tau'') \Im[\zeta_q(\tau) \zeta_q^*(\tau')] \Re[\zeta_q(\tau) \zeta_q^*(\tau'')] \nonumber \\
  &\qquad \qquad \qquad \times \int \frac{\dd^3 k}{(2\pi)^3} \Im[\zeta_k(\tau')\zeta^*_k(\tau'')\zeta_{|\bfq - \bfk|}(\tau')\zeta^*_{|\bfq - \bfk|}(\tau'')] \nonumber \\
  &= \frac{2q^3}{\pi^2} \int^\tau_{\tau_i} \dd \tau' \int^{\tau'}_{\tau_i} \dd \tau'' a^4(\tau') 
  a^4(\tau'') b^3(\tau') b^3(\tau'')V_\tho(\tau') V_\tho(\tau'') \Im[\zeta_q(\tau) \zeta_q^*(\tau')] \Re[\zeta_q(\tau) \zeta_q^*(\tau'')] \nonumber \\
  &\qquad \times  \int \frac{\dd^3 k}{(2\pi)^3} (\Im[\zeta_k(\tau')\zeta^*_k(\tau'')] \Re[\zeta_{|\bfq - \bfk|}(\tau')\zeta^*_{|\bfq - \bfk|}(\tau'')] + (k \leftrightarrow |\bfq - \bfk|) ).
\end{align}
Note that $\mathcal P^a_{\zeta,2\vx}(q,\tau) \propto q^3$ in the superhorizon limit ($|q\tau| \ll 1$) if we focus on the loop contributions from $k \gg q$~\cite{Inomata:2025bqw}.
We also note that $\mathcal P_{\zeta,2\vx}$ corresponds to the contribution explicitly calculated in Ref.~\cite{Kristiano:2024vst} (see also Appendix~\ref{app:comp}).
For later convenience, we here express the expression in the superhorizon limit:
\begin{align}
  \label{eq:p_2vx_lim}
   \lim_{q\to 0}\mathcal P_{\zeta, 2\vx}(q,\tau) &= \frac{4q^3}{\pi^2} \int^\tau_{\tau_i} \dd \tau' \int^{\tau'}_{\tau_i} \dd \tau'' a^4(\tau') 
  a^4(\tau'') b^3(\tau') b^3(\tau'')V_\tho(\tau') V_\tho(\tau'') \Im[\zeta_q(\tau) \zeta_q^*(\tau')] \Re[\zeta_q(\tau) \zeta_q^*(\tau'')] \nonumber \\
  &\qquad \times  \int \frac{\dd^3 k}{(2\pi)^3} \Im[\zeta_k(\tau')\zeta^*_k(\tau'')] \Re[\zeta_k(\tau')\zeta^*_k(\tau'')].
\end{align}

\end{widetext}
Next, we obtain the counterterm contributions.
We first need to determine $V_{c,\fo}$ using the zero-tadpole condition $\expval{\zeta} = 0$.
From Eq.~(\ref{eq:in_in}), the tadpole is given by 
\begin{align}
  \expval{\zeta(\bfx, \tau)} 
  &= -i \int \dd^3 y \int^{\tau}_{\tau_i} \dd \tau' a^4(\tau')b(\tau') [\zeta(\bfx,\tau),\zeta(\bfy,\tau')] \nonumber \\ 
  & \qquad \times \left[ V_{c,\fo}(\tau') + \frac{1}{2} b^2(\tau') V_\tho(\tau') \sigma^2_\zeta(\tau') \right],
\end{align}
where 
\begin{align}
  \int \dd^3 y[\zeta(\bfx,\tau),\zeta(\bfy,\tau')] = 2i \lim_{q\to 0}\Im[\zeta_q(\tau) \zeta^*_q(\tau')].
\end{align}
To satisfy the zero-tadpole condition $\expval{\zeta} = 0$, we tune the counterterm as  
\begin{align}
  V_{c,\fo}(\tau) = - \frac{V_\tho(\tau) b^2(\tau) \sigma^2_\zeta(\tau)}{2} = -\frac{V_\tho(\tau) \sigma^2_{\delta \phi}(\tau)}{2}.
\end{align}
This automatically fixes $V_{c,\so}$:
\begin{align}
  V_{c,\so}(\tau) &= \frac{1}{a(\tau)\dot{\bar \phi}} \frac{\dd V_{c,\fo}(\tau)}{\dd \tau} \nonumber \\ 
  &= -\frac{V_\foo(\tau) b^2(\tau)\sigma^2_\zeta(\tau)}{2}  + \frac{V_\tho(\tau) }{2H a(\tau)b(\tau)} \frac{\dd}{\dd \tau}\sigma^2_{\delta \phi}(\tau),
\end{align}
where $\sigma^2_{\delta \phi}(\tau) \equiv \vev{\delta \phi^2(\bfx,\tau)}$.
We here use ~\cite{Inomata:2024lud,Inomata:2025bqw}\footnote{The right-hand side of this equation can be obtained from the equation around the end of Sec.~IV in Ref.~\cite{Inomata:2025bqw} with $\delta \phi = b\zeta$.}
\begin{align}
    \frac{\dd}{\dd \tau}\sigma^2_{\delta \phi}(\tau) 
  &= H a(\tau) b(\tau)^2 \mathcal P_{\zeta,\tre}\left( k_\ir,\tau \right) \nonumber \\
  &\quad - 4 a(\tau) \int^{\tau}_{\tau_i} \dd \tau' a^4(\tau') b^3(\tau') b^2(\tau) H V_\tho(\tau') \nonumber \\ 
  &\qquad \times \int \frac{\dd^3 k}{(2\pi)^3}\Im[\zeta_k(\tau) \zeta^*_k(\tau')] \Re[\zeta_k(\tau) \zeta^*_k(\tau')].
\end{align}
In the following, we neglect the first term by assuming $\mathcal P_{\zeta,\tre}(k_\ir)$ is negligibly small.

Then, the quadratic counterterm ($\propto V_{c,\so}b^2 \zeta^2$) gives 
\begin{align}
  &\mathcal P_{\zeta,c}(q,\tau) \nonumber \\ 
  &= \frac{2q^3}{\pi^2} \int^{\tau}_{\tau_i} \dd \tau' a^4(\tau') b^2(\tau') V_{c,\so}(\tau') \nonumber \\ 
  &\qquad \qquad \times \Im[\zeta_q(\tau) \zeta^*_q(\tau')] \Re[\zeta_q(\tau) \zeta^*_q(\tau')] \nonumber \\ 
  &= -\frac{q^3}{\pi^2} \int^{\tau}_{\tau_i} \dd \tau' a^4(\tau') \Im[\zeta_q(\tau) \zeta^*_q(\tau')] \Re[\zeta_q(\tau) \zeta^*_q(\tau')] \nonumber \\ 
  &\quad \times \left[ V_\foo b^4(\tau') \sigma^2_\zeta(\tau') \phantom{\int \frac{\dd^3 k}{(2\pi)^3}} \right. \nonumber \\
  &\qquad\quad \left.  + 4 V_\tho(\tau') \int^{\tau'}_{\tau_i} \dd \tau'' a^4(\tau'') b^3(\tau'') b^3(\tau') V_\tho(\tau'')\right. \nonumber  \\ 
  &\quad\left. \times \int \frac{\dd^3 k}{(2\pi)^3}\Im[\zeta_k(\tau') \zeta^*_k(\tau'')] \Re[\zeta_k(\tau') \zeta^*_k(\tau'')] \right].
  \label{eq:p_c}
\end{align}
Combining Eqs.~(\ref{eq:p_1vx}), (\ref{eq:p_2vx_lim}), and (\ref{eq:p_c}), we obtain 
\begin{align}
  \lim_{q \to 0} \mathcal P_\zeta(q,\tau) = \mathcal P_{\zeta,\tre}(q).
\end{align}
The $\mathcal P_{\zeta,1\vx}$ and $\mathcal P_{\zeta,2\vx}$ are canceled by $\mathcal P_{\zeta,c}$.
This means that the curvature perturbations are conserved even at the one-loop level.
We stress again that the contribution of $\mathcal P_{\zeta,c}$ was overlooked in the recent works claiming the non-conservation of superhorizon curvature perturbations at the one-loop level.

\section{Conclusion and discussion}
In this work, we have shown that the counterterms, overlooked in the recent works that claim the non-conservation of superhorizon curvature perturbations, play a crucial role in their conservation.
In particular, we have focused on the one-loop power spectrum in the comoving gauge with the redefined curvature perturbation $\zeta_n$ (we restore $\zeta_n$ in this section to distinguish it from the original curvature $\zeta$). 
Specifically, we have seen that, if we fix the linear counterterm to satisfy the zero-tadpole contribution $\expval{\zeta_n} = 0$, the quadratic counterterm automatically cancels the one-loop power spectrum in the superhorizon limit.
The point is that the linear and the quadratic counterterms are dependent, which is secured by the continuity of the inflaton potential~\cite{Inomata:2025bqw}. 

Although we have used the Hamiltonian in the spatially-flat gauge to obtain that in the comoving gauge with the redefined curvature perturbations, we should in principle be able to obtain the same counterterm contribution by directly redefining the curvature perturbations in the Lagrangian in the comoving gauge.
According to Ref.~\cite{Pimentel:2012tw}, the counterterms in the comoving gauge must be introduced through the following counterterm action,
\begin{align}
  S_{\text{tad,counter}} = \int \dd^4 x \sqrt{-g} \left[g^{00}\delta M^4(t) + \delta \Lambda(t) \right],
\end{align}
where $\delta M^4$ and $\delta \Lambda$ are the counterterms. 
This counterterm contribution was overlooked in works that recently claimed the non-conservation of curvature perturbations. 
The point is that this counterterm action includes nonlinear curvature perturbations by definition, which makes the quadratic counterterms dependent on the linear counterterms, similarly to the relation between $V_{c,\fo}$ and $V_{c,\so}$.
Indeed, in Refs.~\cite{Pimentel:2012tw,Fumagalli:2024jzz}, the authors fix the counterterms by imposing the zero-tadpole condition $\expval{\zeta} = 0$ and see that the quadratic counterterms play a crucial role in the conservation of curvature perturbations.
In particular, in Ref.~\cite{Pimentel:2012tw}, the full expressions of $\delta M^4$ and $\delta \Lambda$ for $\expval{\zeta} = 0$ are obtained.
However, we need to be careful about the fact that $\expval{\zeta} = 0$ is not the same as $\expval{\zeta_n} = 0$ due to the non-linear relation between them.
We leave the direct derivation of the counterterms for $\expval{\zeta_n}=0$ without using the Hamiltonian in the spatially-flat gauge for future work.

On a related note, although we have imposed the zero-tadpole condition $\expval{\zeta_n} = 0$ in this work to demonstrate the importance of the counterterms, the superhorizon curvature perturbations are actually conserved even if $\expval{\zeta_n} \neq 0$ ($\expval{\delta \phi} \neq 0$ in the spatially-flat gauge). In this case, we need to be careful about both the tadpole contributions and the backreaction. 
In Ref.~\cite{Inomata:2025bqw}, we took into account both effects and showed the conservation of the superhorizon curvature perturbations without imposing $\expval{\delta \phi} = 0$ in the spatially-flat gauge.
This result means that the conservation of the superhorizon curvature perturbations is independent of the choice of the counterterms, as long as the dependence between linear and quadratic counterterms is properly considered, regardless of the value of $\expval{\zeta_n}$.

Apart from the one-loop power spectrum, the counterterm contribution can be important when we discuss the one-loop corrections to the bispectrum of the curvature perturbations because $\mathcal H_c \supset V_{c,\tho}b^3 \zeta_n^3/6$ at $\mathcal O(\zeta_n^3)$. The investigation of this topic is left for future work. 

\acknowledgments
\noindent
The author thanks Cheng-Jun Fang, Jacopo Fumagalli, and Jason Kristiano for useful discussions.
The author is grateful to the organizers and the attendees of ``Looping in the Primordial Universe'' workshop at CERN for fruitful discussions.
The author was supported by JSPS Postdoctoral Fellowships for Research Abroad.

\appendix

\begin{widetext}

\section{Comparison with the previous work}
\label{app:comp}

In this appendix, we check that 
\begin{align}
  \label{eq:h_inte_app}
  H_{\inte,3}(\tau) = M_\Pl^2\int \dd^3 x\, d'(\tau) \zeta^3(\bfx,\tau)
\end{align}
leads to the same one-loop contribution from the terms considered in Ref.~\cite{Kristiano:2024vst}:
\begin{align}
  &H_{\inte,\pp}(\tau) = H^\bulk_{\inte,3}(\tau) + H^{4I}_{\inte}(\tau), \\
  &H^\bulk_{\inte,3} \equiv -M_\Pl^2\int \dd^3 x\, d(\tau) (\zeta^3(\bfx,\tau))', \ H^{4I}_{\inte} \equiv \frac{9 M_\Pl^2}{4} \int \dd^3 x\, \frac{d^2(\tau)}{a^2 \epsilon(\tau)} \zeta^4(\bfx,\tau),
  \label{eq:bulk_4i}
\end{align}
where $d(\tau) = a^2(\tau) \epsilon(\tau) \eta'(\tau)/6$.
Eq.~(\ref{eq:h_inte_app}) can be obtained by substituting Eq.~(\ref{eq:v_tho}) into Eq.~(\ref{eq:int_h0_z}).

Specifically, we will see that the two-vertex contribution in this work corresponds to the contributions concretely calculated in Ref.~\cite{Kristiano:2024vst}:
\begin{align}
  &\expval{\zeta_{\bfq}(\tau) \zeta_{\bfq'}(\tau)}_{2\vx}  = 2\, \Re\left[ \int^\tau_{\tau_i}\dd \tau' \int^{\tau'}_{\tau_i}\dd \tau'' \vev{ \left(H_{\text{int},3}(\tau') \zeta^I_{\bfq}(\tau) \zeta^I_{\bfq'}(\tau) - \zeta^I_{\bfq}(\tau) \zeta^I_{\bfq'}(\tau) H_{\inte,3}(\tau') \right) H_{\text{int},3}(\tau'')}\right] \nonumber \\
  &= 2\, \Re\left[ \int^\tau_{\tau_i}\dd \tau' \int^{\tau'}_{\tau_i}\dd \tau'' \vev{ \left(H^\bulk_{\text{int},3}(\tau') \zeta^I_{\bfq}(\tau) \zeta^I_{\bfq'}(\tau) - \zeta^I_{\bfq}(\tau) \zeta^I_{\bfq'}(\tau) H^\bulk_{\inte,3}(\tau') \right) H^\bulk_{\text{int},3}(\tau'')}\right] \nonumber \\ 
  &\quad + 2\, \Im\left[\int^\tau_{\tau_i} \dd \tau'\vev{ \zeta^I_{\bfq}(\tau) \zeta^I_{\bfq'}(\tau) H^{4I}_{\text{int}}(\tau')}\right],
  \label{eq:two_vx_one_loop_app}
\end{align}
where $\expval{\zeta_{\bfq}(\tau) \zeta_{\bfq'}(\tau)}_{2\vx} \equiv (2\pi)^3 \delta(\bfq + \bfq') (2\pi^2/q^3) \mathcal P_{\zeta,2\vx}(q,\tau)$.
The final expression with $H^\bulk_{\inte,3}$ and $H^{4I}_\inte$ corresponds to the contributions calculated in Ref.~\cite{Kristiano:2024vst}.

Note that the following analysis is almost the same as in Ref.~\cite{Fang:2025vhi} except for the last expression, which we will mention later. 
For convenience, we split $\expval{\zeta_{\bfq}(\tau) \zeta_{\bfq'}(\tau)}_{2\vx}$ as 
\begin{align}
  \expval{\zeta_{\bfq}(\tau) \zeta_{\bfq'}(\tau)}_{2\vx} = \expval{\zeta_{\bfq}(\tau) \zeta_{\bfq'}(\tau)}^{(1,1)}_{2\vx} + 2\Re\left[\expval{\zeta_{\bfq}(\tau) \zeta_{\bfq'}(\tau)}^{(0,2)}_{2\vx} \right],
\end{align}
where 
\begin{align}
  \expval{\zeta_{\bfq}(\tau) \zeta_{\bfq'}(\tau)}^{(1,1)}_{2\vx} &= \int^\tau_{\tau_i}\dd \tau' \int^{\tau}_{\tau_i}\dd \tau'' \vev{ H_{\text{int},3}(\tau') \zeta^I_{\bfq}(\tau) \zeta^I_{\bfq'}(\tau) H_{\text{int},3}(\tau'')}, \\ 
  \expval{\zeta_{\bfq}(\tau) \zeta_{\bfq'}(\tau)}^{(0,2)}_{2\vx} &=  -\int^\tau_{\tau_i}\dd \tau' \int^{\tau'}_{\tau_i}\dd \tau'' \vev{ \zeta^I_{\bfq}(\tau) \zeta^I_{\bfq'}(\tau) H_{\inte,3}(\tau') H_{\text{int},3}(\tau'')}.
\end{align}
After some calculation, we obtain
\begin{align}
  \label{eq:zeta_2vx_11}
  \expval{\zeta_{\bfq}(\tau) \zeta_{\bfq'}(\tau)}^{(1,1)}_{2\vx} &= M_\Pl^4\int \dd^3 x \int \dd^3 y \int^{\tau}_{\tau_i} \dd \tau' \int^{\tau}_{\tau_i} \dd \tau'' d'(\tau') d'(\tau'') \vev{(\zeta^I(\bfx,\tau'))^3 \zeta^I_{\bfq}(\tau) \zeta^I_{\bfq'}(\tau) (\zeta^I(\bfy,\tau''))^3} \nonumber \\
  &= M_\Pl^4 \left[ \int \dd^3 x \int \dd^3 y d^2(\tau) \vev{(\zeta^I(\bfx,\tau))^3 \zeta^I_{\bfq}(\tau) \zeta^I_{\bfq'}(\tau) (\zeta^I(\bfy,\tau))^3} \right. \nonumber \\
  &\quad - 2\int \dd^3 x \int \dd^3 y \int^{\tau}_{\tau_i} \dd \tau' d(\tau') d(\tau) \Re[\vev{ ((\zeta^I(\bfx,\tau'))^3)' \zeta^I_{\bfq}(\tau) \zeta^I_{\bfq'}(\tau) (\zeta^I(\bfy,\tau))^3}] \nonumber \\ 
  &\quad \left. + \int \dd^3 x \int \dd^3 y \int^{\tau}_{\tau_i} \dd \tau' \int^{\tau}_{\tau_i} \dd \tau'' d(\tau') d(\tau'') \vev{ ((\zeta^I(\bfx,\tau'))^3)' \zeta^I_{\bfq}(\tau) \zeta^I_{\bfq'}(\tau) ((\zeta^I(\bfy,\tau''))^3)'} \right],
\end{align}
\begin{align}
  \label{eq:zeta_2vx_02}
\expval{\zeta_{\bfq}(\tau) \zeta_{\bfq'}(\tau)}^{(0,2)}_{2\vx} &= - M_\Pl^4\int \dd^3 x \int \dd^3 y \int^{\tau}_{\tau_i} \dd \tau' \int^{\tau'}_{\tau_i} \dd \tau'' d'(\tau') d'(\tau'') \vev{\zeta^I_{\bfq}(\tau) \zeta^I_{\bfq'}(\tau) (\zeta^I(\bfx,\tau'))^3 (\zeta^I(\bfy,\tau''))^3} \nonumber \\
&= M_\Pl^4 \left[- \frac{1}{2} \int \dd^3 x \int \dd^3 y d^2(\tau) \vev{\zeta^I_{\bfq}(\tau) \zeta^I_{\bfq'}(\tau) (\zeta^I(\bfx,\tau))^3 (\zeta^I(\bfy,\tau))^3} \right.\nonumber \\ 
&\quad + \int \dd^3 x \int \dd^3 y \int^{\tau}_{\tau_i} \dd \tau' d(\tau) d(\tau') \vev{\zeta^I_{\bfq}(\tau) \zeta^I_{\bfq'}(\tau) (\zeta^I(\bfx,\tau))^3 ((\zeta^I(\bfy,\tau'))^3)'} \nonumber \\ 
&\quad + \frac{1}{2} \int \dd^3 x \int \dd^3 y \int^{\tau}_{\tau_i} \dd \tau' d^2(\tau') \vev{\zeta^I_{\bfq}(\tau) \zeta^I_{\bfq'}(\tau) \left[ ((\zeta^I(\bfx,\tau'))^3)', \zeta^3(\bfy,\tau') \right]} \nonumber \\ 
&\quad \left.- \int \dd^3 x \int \dd^3 y \int^{\tau}_{\tau_i} \dd \tau' \int^{\tau'}_{\tau_i} \dd \tau'' d(\tau') d(\tau'') \vev{\zeta^I_{\bfq}(\tau) \zeta^I_{\bfq'}(\tau) ((\zeta^I(\bfx,\tau'))^3)' ((\zeta^I(\bfy,\tau''))^3)'} \right].
\end{align}
The first and the second terms in the last equalities of Eqs.~(\ref{eq:zeta_2vx_11}) and (\ref{eq:zeta_2vx_02}) cancel. 
From these, we can obtain Eq.~(\ref{eq:two_vx_one_loop_app}) as 
\begin{align}
    \expval{\zeta_{\bfq}(\tau) \zeta_{\bfq'}(\tau)}_{2\vx} &=  M_\Pl^4 \left\{\int \dd^3 x \int \dd^3 y \int^{\tau}_{\tau_i} \dd \tau' \int^{\tau}_{\tau_i} \dd \tau'' d(\tau') d(\tau'') \vev{ ((\zeta^I(\bfx,\tau'))^3)' \zeta^I_{\bfq}(\tau) \zeta^I_{\bfq'}(\tau) ((\zeta^I(\bfy,\tau''))^3)'} \right.\nonumber \\ 
    &\quad 
    - 2 \Re\left[ \int \dd^3 x \int \dd^3 y \int^{\tau}_{\tau_i} \dd \tau' \int^{\tau'}_{\tau_i} \dd \tau'' d(\tau') d(\tau'') \vev{\zeta^I_{\bfq}(\tau) \zeta^I_{\bfq'}(\tau) ((\zeta^I(\bfx,\tau'))^3)' ((\zeta^I(\bfy,\tau''))^3)'} \right] \nonumber \\
    &\quad \left.+ \Re\left[ \int \dd^3 x \int \dd^3 y \int^{\tau}_{\tau_i} \dd \tau' d^2(\tau') \vev{\zeta^I_{\bfq}(\tau) \zeta^I_{\bfq'}(\tau) \left[ ((\zeta^I(\bfx,\tau'))^3)', \zeta^3(\bfy,\tau') \right]} \right] \right\} \nonumber \\
    &=  M_\Pl^4\int \dd^3 x \int \dd^3 y \int^{\tau}_{\tau_i} \dd \tau' \int^{\tau}_{\tau_i} \dd \tau'' d(\tau') d(\tau'') \vev{ ((\zeta^I(\bfx,\tau'))^3)' \zeta^I_{\bfq}(\tau) \zeta^I_{\bfq'}(\tau) ((\zeta^I(\bfy,\tau''))^3)'} \nonumber \\ 
    &\quad 
    - 2 M_\Pl^4\Re\left[ \int \dd^3 x \int \dd^3 y \int^{\tau}_{\tau_i} \dd \tau' \int^{\tau'}_{\tau_i} \dd \tau'' d(\tau') d(\tau'') \vev{\zeta^I_{\bfq}(\tau) \zeta^I_{\bfq'}(\tau) ((\zeta^I(\bfx,\tau'))^3)' ((\zeta^I(\bfy,\tau''))^3)'} \right] \nonumber \\
    &\quad + \frac{9M_\Pl^2}{2} \int \dd^3 x \int^{\tau}_{\tau_i} \dd \tau' \frac{d^2(\tau')}{a^2(\tau') \epsilon(\tau') } \Im[\vev{\zeta^I_{\bfq}(\tau) \zeta^I_{\bfq'}(\tau) \zeta^4(\bfx,\tau')}] \nonumber \\ 
    &= 2\, \Re\left[ \int^\tau_{\tau_i}\dd \tau' \int^{\tau'}_{\tau_i}\dd \tau'' \vev{ \left(H^\bulk_{\text{int},3}(\tau') \zeta^I_{\bfq}(\tau) \zeta^I_{\bfq'}(\tau) - \zeta^I_{\bfq}(\tau) \zeta^I_{\bfq'}(\tau) H^\bulk_{\inte,3}(\tau') \right) H^\bulk_{\text{int},3}(\tau'')}\right] \nonumber \\ 
  &\quad + 2\, \Im\left[\int^\tau_{\tau_i} \dd \tau'\vev{ \zeta^I_{\bfq}(\tau) \zeta^I_{\bfq'}(\tau) H^{4I}_{\text{int}}(\tau')}\right],
\end{align}
where we have used the commutation relation $[\zeta'(\bfx,\tau), \zeta(\bfy,\tau)] = -i \delta(\bfx- \bfy)/(2a^2(\tau)\epsilon(\tau) M_\Pl^2)$ in the second equality. 
The expression is the same as Ref.~\cite{Fang:2025vhi} except for the factor of the last terms in the second and the third equalities, while the expression in the first equality is exactly the same.

\end{widetext}

\small
\bibliographystyle{apsrev4-1}
\bibliography{draft_sh_one_loop_in_comoving_gauge}

\end{document}